\documentclass[useAMS,usenatbib,preprint]{aastex}
\usepackage{amsmath,amssymb}
\usepackage{natbib}
\usepackage{url}
\usepackage{graphicx}
\usepackage{times}
\usepackage{ulem}
\usepackage{colordvi}

\def\ifnote{\iftrue}

\begin{document}
\title{GRB Light Curves in the Relativistic Turbulence and Relativistic Sub-Jets Models}
\author{Ayah Lazar$^{1,2}$, Ehud Nakar$^{3,4}$ and Tsvi Piran$^{1}$}
\affil{1. The Racah Institute of Physics, Hebrew University, Jerusalem 91904, Israel
\\ 2. Department of Geophysics and Planetary Sciences, Tel Aviv University, Tel Aviv  69978 Israel\\
3. Raymond and Beverly Sackler School of Physics \&
Astronomy, Tel Aviv University, Tel Aviv 69978, Israel
\\ 4. Theoretical Astrophysics, Caltech, Pasadena, CA 91125, USA}



\begin{abstract}
Randomly oriented relativistic
emitters in a relativistically expanding shell provides an
alternative to internal shocks as a mechanism for producing GRBs'
variable light curves with efficient conversion of energy to
radiation. In this model the relativistic outflow is broken into
small emitters moving relativistically in the outflow's rest frame.
Variability arises because an observer sees an emitter only when its
velocity points towards him so that only a small fraction of the
emitters are seen by a given observer.
Models with  significant relativistic random
motions require  converting and maintaining a large fraction of the
overall energy into these motions. While it is not clear how this is
achieved, we explore here, using two toy models, the constraints on
parameters required to produce  light curves comparable to the
observations. We find that a tight relation between the size of the
emitters and the bulk and random
Lorentz factors is needed and that the  random Lorentz factor determines the
variability. While both models successfully
produce the observed variability there are several inconsistencies
with other properties of the light curves. Most of which, but not
all, might be resolved if the central engine is active for a long
time producing a number of  shells, resembling to some extent the
internal shocks model.

\end{abstract}

\keywords{gamma-rays: bursts; turbulence; }

\section{Introduction}

GRB's  temporal variability  played a major
role in the  understanding how  GRBs operate. Standard
external shocks, in which the external medium slows the relativistic ejecta, cannot produce efficiently  variable
light curves \citep{SP1997}.
While internal shocks  resolve the variability and agree with other properties of GRB light curves
(e.g. \citealt{NakarPiran02b,Ramirez2000})
they suffers from several drawbacks. First and foremost is their  low efficiency
(\citealt{KPS97,DM98}  see however,
\citealt{KPS97,KS2001,B2000}). This
is particularly troublesome in view of the high efficiency implied
from comparison of the prompt $\gamma$-rays luminosity and the
kinetic energy that remains in  the outflow. Detailed models for the emission
mechanisms of the prompt $\gamma$-rays pose other problems \citep{KM08}.

External shocks can produce highly variable light curves if the
outflow is slowed down by small external clumps. Each clump
producing a short pulse. However, this process will inevitably be
 inefficient \citep{SP1997} as the overall covering
factor of the emitting regions is $\delta t/T$ ($\delta t$  and $T$ are
the pulses' and the burst's durations).  Observed values of $\delta t /T$ are typically
$\sim 0.01$ and in  can be as low as $10^{-4}$
\citep{NakarPiran02a}.

Lyutikov \& Blandford (2002, 2003)  (see also \citealt{LazarPiran-1}
[L05], \citealt{Lyutikov06}, \citealt{NarayanKumar2008}) proposed
that variability can be recovered while maintaining high efficiently
it the shell that moves with a bulk Lorentz factor $\Gamma$
contains emitting clumps (see fig. \ref{fig:kinemticmodel}) that
move with random macroscopic relativistic velocities (with a Lorentz
factor $\gamma'$). A clump is observed only when its radiation cone
(with an opening angle of the order of $1/\Gamma\gamma'$ in the lab
frame) points towards the observer. The filling factor of the clumps
may be unity, recovering high efficiency. However, as only a small
fraction of the clumps are observed at any given time, the light
curve can show  rapid variability. The overall duration of the
burst, is the larger between the angular time and the shell's light
crossing time ($\max\{R/c \Gamma^2,\Delta/c\}$), where $R$ and
$\Delta$ are the shell's radius and width respectively and $c$ is the
light speed,  allowing for emission radii much larger than $\delta t \Gamma^2$, required
in the internal shocks model.
 The temporal variability is then dictated by the random
Lorentz factor, $\gamma'$, reflecting the activity of the emitting
region and not those of the inner engine.

While it is unclear how macroscopic random relativistic
motion can be generated, we assume  that it does
and examine, using two simplified toy models that includes the
essential ingredients, the conditions under which  the  temporal
features of the observed light curve can be produced  (see L05). We
describe our first  toy model, which we call here relativistic
turbulence, and derive analytic constraints and numerical
light curves in \S \ref{sec:The-Relativistic-Turbulence}. In  \S
\ref{sec:Fundamental-emitters} we consider a second toy model
proposed by Lyutikov (2006) that is based on
sub-jets and compare it with the first one. We summarize
the results and compare both models with observations in \S \ref{sec:
conclusions}.

\section{Relativistic Turbulence} \label{sec:The-Relativistic-Turbulence}

\begin{figure}[h]
\includegraphics[width=0.5\textwidth]{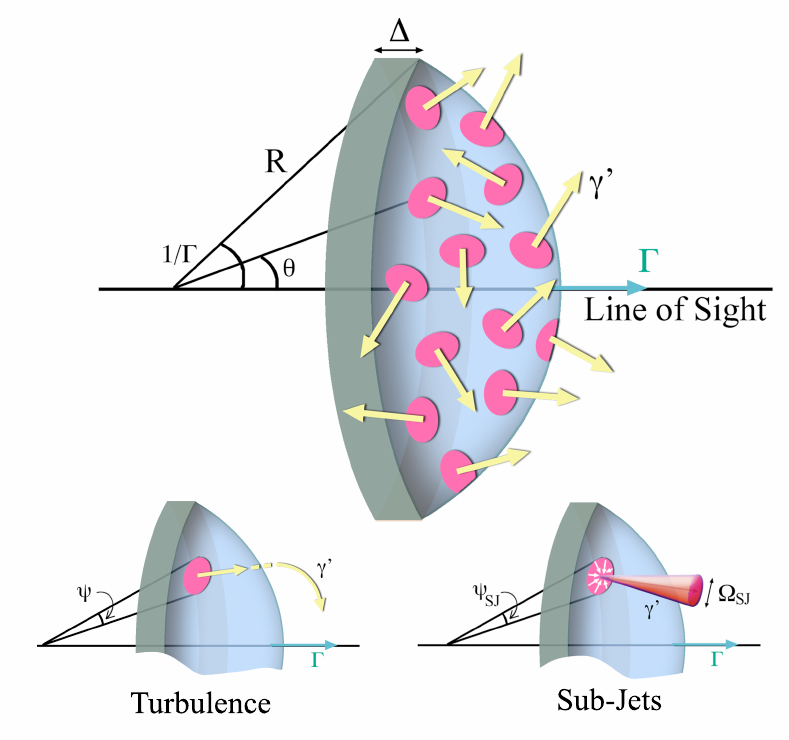}
\caption{The basic kinematic model -- relativistically
expanding shell with an ensemble of emitters,
which are moving at random relativistic velocities within the shell.
The inserts describe the geometrical details and definitions of the two alternative models. Left: relativistic turbulence - emitters of size
$R\psi$ move relativistically;  Right:  sub-jets - energy from  regions of size $R\psi_{SJ}$ is extracted into relativistic  jets. }
\label{fig:kinemticmodel}
\end{figure}

Our (L05) kinematic toy model for relativistic turbulence  considers
a shell which is divided into discrete randomly distributed emitters
that have randomly oriented relativistic velocities. The emitters
change their direction of motion continuously, as expected in a
turbulent medium. Since the emitters exhibit a coherent macroscopic
motion, we require that each emitter is causally connected, and that
it changes its direction on times longer than the causal time scale
but shorter than the shell crossing time. The
length scale of the emitter in its own rest frame, $l''$, is assumed
to be similar in all three dimensions, and it emits isotropically in
this frame. We define a dimensionless scale $\psi \equiv l''/R$,
which is the angular scale of an emitter that points towards the
observer (see inset of fig \ref{fig:kinemticmodel}). The emitters
radiate as the shell moves from $R_0$ to $2R_0$. Due to the
turbulent motion the positions and directions of the emitters change
with time. We model this by a set of successive shells between $R_0$
and $2R_0$. Each new shell is constructed with randomly distributed
emitters, representing the random changes in direction of the
turbulent motion. The time difference between two shells is,
$\tau'$, the time it takes for the emitters to turn an angle of
${\gamma'}^{-1}$ (in the shell frame).

Note that there are three frames: The lab frame; the shell's frame,
denoted by a prime, which is boosted radially with a Lorentz factor
$\Gamma$ relative to the lab; and the frame of each emitter, denoted
by two primes, which is boosted by (randomly oriented) $\gamma'$
relative to the shell frame. The observer is, of course, at rest
relative to the lab frame. However, the observer time, namely the
arrival time of photons (denoted $t$) differs from the lab time by
the usual time of flight arguments \citep{Rybicki}.

The Doppler shift from an emitter is:
\begin{equation}
\Lambda =[{\gamma(1-\beta\cdot \cos\alpha)}]^{-1}  ,
\label{eq: Doppler shift}
\end{equation}
where  $\gamma$, $\beta$ and   $\alpha$ are the Lorentz factor,  velocity  and  the angle between the velocity and the line to the observer (both in the lab
frame).
The flux  that  reaches the observer  from this emitter is:
\begin{equation}
F_{\nu}  =  \int I''_{\nu''}  \Lambda^{3}  d\Omega_i \approx
I''_{\nu''}  \Lambda^{3}\frac{\psi^{2}R^{2}}{D^{2}},
\label{eq: Energy}
\end{equation}
where $ I''_{\nu''} $ is the  specific intensity and the second relation holds for a small enough emitter ($D$
is the distance to the observer). An  implicit
K correction arises from the difference between the $\nu$ and $\nu''$.

Define $\theta$ as the angle between the line
that connects the origin and the emitter and the line that connects
the origin and the observer (see inset of fig
\ref{fig:kinemticmodel}). The maximal Doppler boost, $\Lambda_{max}
= {4\gamma'\Gamma}$, is obtained for emitter at $\theta=0$ that
moves along the line of sight. The flux decreases like $\Lambda^3$ plus a K correction. Therefore, we
consider emitters only if
$\Lambda_i > \Lambda_{max}/2$. What is then the probability that
an emitter at angle $\theta$
is observed, $dP/d\theta \equiv S(\theta,\Gamma,\gamma')$? At $\theta=0$ it is
$S(0,\Gamma,\gamma') \approx {1}/{4\gamma'^{2}}$
while $S({1}/{\Gamma},\Gamma,\gamma')=0$. This
suggests  that $S$  scales as $S(\theta,\Gamma,\gamma')  =
{\gamma'}^{-2} \tilde S({\theta}{\Gamma})$ (L05),
implying that the average probability that an
emitter will be visible from an arbitrary position  on the shell is:
\begin{equation}
P(\Gamma,\gamma')\approx  {1 \over 2 (\Gamma \gamma')^2 } \int_{0}^{1}\tilde S(\theta\Gamma)(\theta \Gamma) d(\theta \Gamma) \approx   \frac{0.3}{4 \pi (\Gamma \gamma')^{2}} .
\label{eq:probability}
\end{equation}
The factor 0.3 was evaluated numerically (L05) and  can be ignored
at the accuracy level of our discussion.

The arrival time  from an emitter
at $R, \theta$ is:
\begin{equation}  \label{eq:Arrival-Time}
T=\frac{R-R_{0}}{2c\Gamma^{2}}+\frac{R\theta^{2}}{2c}+\frac{x}{c},
\end{equation}
where $x(<\Delta)$ is the distance of the emitter from the front of
the shell. As the last photons will arrive from $2R_0$, an angle of
${1}/{\Gamma}$ and $x=\Delta$, the overall duration of the burst
will be a function only of $\Delta$ and $\Gamma$ (and not
 $\gamma'$):
\begin{equation}
T\approx\frac{R_{0}d}{c\Gamma^{2}}. \label{eq:Burst-Duration}
\end{equation}
where we define $ d \equiv {\Delta\Gamma^2}/{R}$.
As the shell is expected to expand relativistically in its own frame\footnote{Note that for a hydrodynamic external shock $d \lesssim 1$ \cite{SP1997} but this might not be relevant here.}
$d \gtrsim 1$. For $d>1$ the shell's width, as well as $T$, are determined by the engine activity
time while for $d=1$ they  don't.

The duration of a pulse arriving from a single emitter  is the
longest of the three following time scales:

(i) The duration over which the emitter points towards the observer,
namely the duration over which the direction of motion varies by an
angle $1/\Gamma\gamma'$ in the lab frame ($1/\gamma'$ in the shell's
frame). As the  emitter is confined to the shell it should make at
least a $\pi/2$ turn during $\Delta'/c$, implying that the time to
turn by $1/\gamma'$(shell's frame), $\tau'$, is shorter than
$\Delta'/c \gamma'$. Causality puts a lower limit on $\tau'$  of $R
\psi / c $. Therefore:
\begin{equation}\label{eq: tau'}
R \psi /c \le \tau' \le \Delta' /c \gamma'.
\end{equation}
In the observer's frame, this translates to:
\begin{equation}\label{eq: tau}
R \psi /\Gamma \gamma'^2 c \le \tau \le \Delta /c \gamma'^3.
\end{equation}

(ii) The emitter's light crossing time in the lab frame (in the
direction along the line of sight). For an emitter moving towards the observer this time is ${R\psi}/{\gamma'\Gamma}$.
(iii) The angular time scale -- At the largest possible angle, where
the emitter is still visible by the observer, ${1}/{\gamma'\Gamma}$,
the time difference between the first and the last photon would be
$\frac{1}{c}R\psi \sin\left({1}/{\gamma'\Gamma}\right)\thickapprox
{R\psi}/{c\gamma'\Gamma}$. Overall (ii) and

(iii) are of the same
order and much larger than (i). Thus:
\begin{equation} \label{eq:Peak-Duration}
\delta t \approx {R\psi}/{c\gamma'\Gamma} .
\end{equation}

Using Eqs.  \ref{eq:Burst-Duration} and \ref{eq:Peak-Duration} we express,  $N_{p}$, the
number of (possibly overlapping)  pulses expected in a burst:
\begin{equation}
N_{p}\equiv n_p\frac{T}{\delta
t}=n_p\frac{d\gamma'}{\psi\Gamma}\label{eq:Num-of-Pulses}
\end{equation}
where $n_p$  is the occupation number of pulses at any given
observer time (i.e., $n_p\gg 1$ implies many overlapping pulses
while $n_p\ll1$ implies long quiescent periods between isolated
pulses).

The number of emitters is ${4 \pi R^{2}\Delta'}/
{(R\psi)^{3}}={ 4 \pi \Delta \Gamma}/ {R \psi^{3}}$. The emitters obtain new random directions (which differ by
more than  $1/\gamma'$, in the shell's frame, than the previous
ones) after a time $\tau'$. Thus,  the total number of independent
emitters, $N_{tot}$, is larger by the factor $R/(c\Gamma\tau')$, the
ratio of  the total duration over which the radius doubles and
$\tau'$. Finally we introduce a filling factor $f \le 1$ allowing
for the possibility that not all emitters are active all the time or
that space is not fully covered by emitters ($f \ll 1$ is
strongly disfavored as the efficiency is always smaller than $f$).
Overall we find:
\begin{equation} \label{eq:Num-of-Emitters}
N_{tot}=\frac{ 4 \pi  f}{ \psi^{3}}\frac{d}
{\Gamma^2}\frac{R}{c\tau'} .
\end{equation}
The condition $N_P =  P N_{tot}$ yields:
\begin{equation}
n_p = \frac {f d}{\gamma'^3 \Gamma^3 \psi^{2}} \frac{R}{c\tau'},
\end{equation}
and using \ref{eq: tau}:
\begin{equation}
\frac{f}{d(\gamma' \Gamma \psi)^2} \le n_p \le \frac{f}{(\gamma'
\Gamma \psi)^3} . \label{eq:xx}
\end{equation}

We demand $n_p\approx 1$ since many overlapping
pulses reduce the observed variability, whereas very frequent long
quiescent times between  pulses are not observed.
If  the shell is in the freely expanding phase (i.e.,
$d \approx 1$)  $n_p$ will be of order unity if:
\begin{equation}
\psi\thickapprox
f^{1/k}\frac{1}{\gamma'\Gamma},\label{eq:Emitter-size}
\end{equation}
where $k$ is between 2 and 3. \cite{NarayanKumar2008} have pointed
out that $\psi =1/\gamma' \Gamma$  if one requires that the emitters
are of the maximal causally allowed size.  Note that $n_p$ depends
quite sensitively on $\gamma' \Gamma \psi$ and it increases rapidly
if $\psi$ is smaller than $1/\gamma' \Gamma$. This implies, for
example, that a significant number of small eddies,
which may arise in a turbulent cascade, may be
problematic.  Using the relations \ref{eq:Emitter-size} and
\ref{eq:Num-of-Pulses}, and assuming the causal limit for $\tau'$:
\begin{equation} \label{eq: gamma'}
\gamma' \approx \left( \frac{f}{n_p}\right)^{1/6} \sqrt{\frac{T}{d
\delta t}},
\end{equation}
leading to $\gamma' \approx
10/\sqrt{d} $ for typical values of $T/\delta t$.
Note that while the model determines $\gamma'$ it does not constrain
$\Gamma$ and $R$.

\begin{figure}[h]
\includegraphics[width=0.5\textwidth]{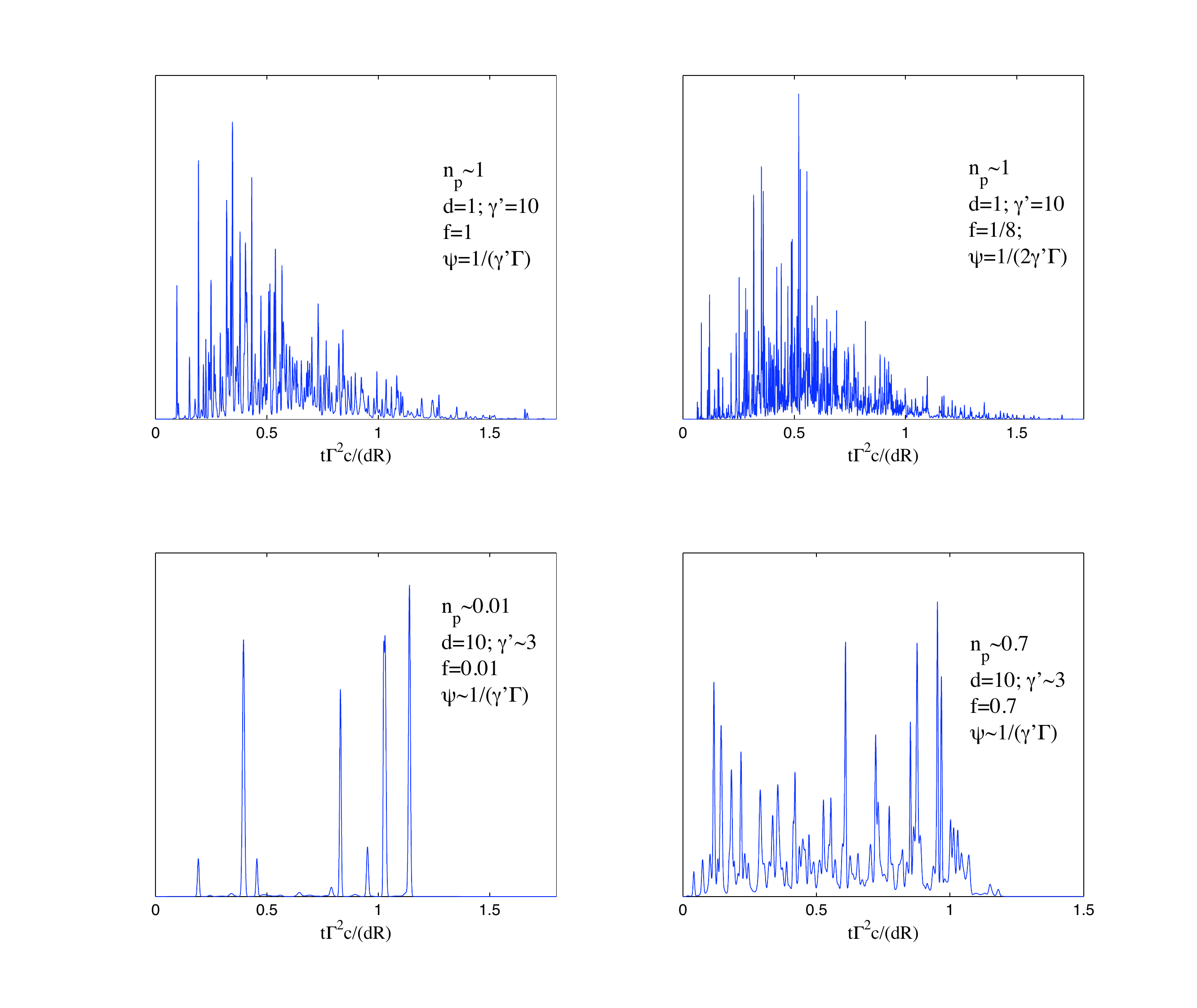}
\caption{\label{cap:lightcurvesf} Numerical monte Carlo simulations of light curves for different combinations of
parameters (shown on each frame; the scalings used eliminate the dependence on $R$ and $\Gamma$).
The flux of each pulse
is calculated assuming that the radiation efficiency is constant per
unit mass for all emitters in their rest frame, namely that
$I''_{\nu''}\propto(\psi R)^{-3}$, therefore $ F_{\nu} \propto
\Lambda^{3}/{(\psi R)}$. We approximate each pulse as a Gaussian
with the above parameters and we sum over all contributions to construct
a light curve. In both upper frames $n_p = 1$ and $d=1$.
In the top left, the emitters are as large as causality allows while on the top right they are
1/2 of this value and the filling factor is lowered to compensate. In both the overall envelope is seen clearly.
The sparsity of pulses is apparent when $n_p=0.01$ (bottom left) and the "straightening out" of the envelope
is clearly seen  (bottom two panels) when $d=10$.
}\end{figure}

Fig. \ref{cap:lightcurvesf} depicts simulated light curves (L05) for
four choices of parameters. The two upper panels have $n_p=1$ with
different  emitter sizes. Both light curves are highly variable and
densely filled with non-overlapping pulses. However, as $d=1$, the
underlying overall envelope of the pulses is seen. As the emitters
are smaller on the right panel it has more pulses than the left one.
The envelope is observed since only a small fraction of the
volume and hence fewer pulses are seen early on. Similarly at $t>
(d+1) R_0/2 c \Gamma^2$ pulses from small $\theta$ values are not
seen, implying that only lower amplitude pulses
(on average) are observed during the last $T/(d+1)$ of the burst. The
envelope is stretched on bottom panels where  $d=10$. The lower left
panel depicts a very low $n_p$ with a rather sparse light curve. The
lower right panel depicts  a light curve of a wide shell and
$n_p=0.7$, which is rather similar to observed bursts.  For $n_p \gg
1$ ( not shown ) the pulses are overlapping and all variability is erased, leaving
only the envelope.

\section{Relativistic Sub-Jets}
\label{sec:Fundamental-emitters} Motivated by  reconnection in
highly magnetized outflow Lyutikov (2006) considers a  model in
which relativistic sub-jets (SJs) are accelerated to a Lorentz
factor $\gamma'$ by dissipation of the bulk energy in many different
"mini-engines" within the relativistically expanding shell. These
"mini-engines" or acceleration sites correspond to reconnection
sites within the magnetized flow (e.g.,
\citealt{LyutikovBlackman01}). The "mini-engines"  are at rest in
the shell frame. Each mini-engine operates for a time $t'_{SJ}$. The
directions of the accelerated sub-jets are random in the shell frame
but the opening angle and direction of each is constant while its
mini-engine is active. 

The sub-jet extracts energy from a region of size $l' = \delta
t'_{SJ} \hat \beta c$, where $\hat \beta c $ is the speed of
extraction of energy from the surrounding region (relativistic
reconnection suggests $\hat \beta \approx 0.1$,
\citealt{Lyubarsky05}).
The observed duration is:
\begin{equation}
\delta t_{SJ} = \frac{t'_{SJ}}{\Gamma} \approx \frac{R \psi_{SJ} }{
\hat \beta c \Gamma} , \label{Ldt}
\end{equation}
where we define the dimensionless parameter $\psi_{SJ}
\equiv l'/R$.
Slightly generalizing \cite{Lyutikov06} we write the probability to
observe an emitter  as $\phi^2/4 \pi \Gamma^2$, where $\phi =
\max(\sqrt{\Omega_{SJ}}, 1/\gamma')$ and $\Omega_{SJ}$ is the
sub-jet opening solid angle. Following the notation  of
\S\ref{sec:The-Relativistic-Turbulence},  we define, $f$, the
filling factor of regions from which energy is extracted into the
sub-jets ($  N_{tot} \equiv f 4\pi\Delta'R^2/l'^3$). The occupation number of observed pulses is:
\begin{equation}
n_{p_{SJ}} = \frac {f  \phi^2 }{ \hat \beta \psi_{SJ}^2
\Gamma^2 }. \label{Lnp}
\end{equation}
The condition $n_{p_{SJ}} \approx 1 $ yields:
\begin{equation}
\psi_{SJ}  \approx  \frac{\sqrt f \phi}{ \hat \beta^{1/2}\Gamma}, \label{psi_sj}
\end{equation}
and
\begin{equation}
\frac{T}{\delta t} \approx \frac{\hat \beta^{3/2} d}{\phi \sqrt f}
\leq \gamma' \frac{\hat \beta^{3/2} d}{\sqrt f}. \label{Lconst}
\end{equation}
This implies that an efficient ($
f \sim 1$) highly variable burst requires either a  large  $\gamma'$ or a wide shell (for $T/\delta t \sim 100$, $\gamma' d \sim 100\hat \beta^{-3/2}$.).

This constant direction of the emitters and  the fact that  causality in the shell's frame determines
the sub-jet size,  $c
\delta t'$,  are the
main kinematic differences between the sub-jet and the turbulence
model (in which the emitter's direction varies and causality in the emitter's
frame determine its 
size, $l''$).
For the same $n_p$ and $\delta t/T$ the two models give similar light curves. 
In particular, an overall (rising and falling) envelope
for the light curve is expected in the sub-jet model as well.

\section{Discussion and Conclusions}\label{sec: conclusions}

We have derived conditions on the parameters of relativistic
random emitters needed for producing variable
GRB light curves. This is characterized by   $n_p \approx
1$ which ensures that typical  pulses don't
overlap and are not too sparse either. Our numerical simulations
show that for $0.03< n_p < 3 $ one obtains
light curves that resemble observed GRBs (see fig.
\ref{cap:lightcurvesf}). The resulting light curves do not change
qualitatively when we introduce a distribution of turbulent Lorentz
factors and  sizes.

Causality  suggests, for relativistic turbulence,
that the relation $\psi = 1/\Gamma \gamma' $ between the
angular size of the emitters, $\psi$, and the turbulent and the bulk
Lorentz factors holds naturally \citep{NarayanKumar2008}. But, this
condition holds when the turbulent eddies are of the maximal
possible size and may be broken by cascade to lower scales.  The condition,
 $\psi_{SJ} =\sqrt{f} \phi /\hat \beta^{3/2}\Gamma$ arises in  the sub-jets model.
For high efficiency,
negligible sub-jet opening angle \and assuming $\hat \beta
\sim 1$ this reduces to $\psi_{SJ} \sim 1 /\Gamma \gamma'$ or
$t'_{SJ} \approx R /c \Gamma \gamma'$. While this  is
similar to the one obtained in the turbulent model, here
there is no apparent physical motivation for proportionality between
$t'_{SJ}$ and $1/\gamma'$ and this  requires an ad hoc fine
tuning.

In both models the light curves arising from a single expanding shell
with $d \approx 1$ shows a rising and falling underlying envelope.
Furthermore, a single shell cannot produce bursts which depict
long  quiescent periods.
The envelope can be erased if $d \gg 1$, while quiescent periods
require an outflow of several shells (where naturally $d \gg 1$).
These solutions become marginal in the turbulent model if $\tau'$ is
determined by causality, since $\gamma' \gg 1$ requires $d \lesssim
10$    (see Eq. \ref{eq: gamma'}). The sub-jet model, however, may
favor $d \gg 1$ as it reduces the required value of $\gamma'$.

It seems that with proper conditions (and rather reasonable in the
case of the relativistic turbulence) these models can produce
(efficiently) the observed highly variable GRB light curves.  We
turn now to several shortcomings. First and foremost is the question
how such macroscopic relativistic motions can be generated and
sustained. One needs to convert $\sim(1-1/\gamma') f$ of the initial
total energy to the kinetic energy of the emitters and further
dissipation in the emitters' frame is needed to generate the
radiation. Additional questions involve the shape and other
properties of individual pulses versus those seen in observed
pulses: \hfill\break\noindent (i)  GRBs show a clear difference
between the fast rise and the slow decline of individual pulses
\citep{Norris1996}.  The light curve of an individual pulse results
from a combination of the motion of the emitter, its orientation
relative to the observer, its width as well as intrinsic
inhomogeneities within the emitter. In the relativistic turbulence
model the emitter was radiating long before its velocity pointed
towards the observer and it continues to emit long after it moves
away from the observer. There is no reason (on average) for a
difference between the rising and falling phases of an individual
pulse\footnote{Note that systematic variation of the emitter
properties on a time scale of $\tau'$ will result in a strong
signature differentiating between early and late phases of the
overall light curve, which is not observed. On the other hand
non-systematic variations (e.g., deceleration and acceleration) are
expected to result in similar affects on the temporal structure of
rising and decaying parts of pulses.}. This is not a problem in the
sub-jets model in which the onset  of the pulse corresponds to the
beginning of the activity of the emitter. \hfill\break\noindent (ii)
The temporal structure of the first and second halves of GRB light
curves are similar \citep{Ramirez2000}. The light curves produced in
the two models have an overall envelope that favors stronger pulses earlier
and weaker ones later. This might be resolved by a combination of several emitting
shells or with very wide shells, but here  fine tuning is
required in the turbulent model in order to keep $\gamma' \gg 1$.
\hfill\break\noindent (iii) Weaker and denser pulses (arriving from
emitters not moving directly towards the observer) continues  at
$t>T$ producing the typical envelope of high latitude emission
\citep{KP00}. This is consistent with  rapid declines seen in some
the early afterglows. However in many cases the decline is faster. In
the  internal shocks model this is attributed to the
dominant contribution of a late  pulse, that
shifts the zero point of the time. Such an option does not arise here unless, once more, we
allow for several shells or a single wide shell.
\hfill\break\noindent (iv) The duration of an observed pulse is
correlated with the preceding interval \citep{NakarPiran02b,Quilligan2002}.  There is no reason that
such correlation should appear in both models. \hfill\break\noindent
(v)  These models  predict a Doppler induced correlation
between the intensity and $E_{peak}$. While stronger peaks are
typically harder, it is not clear whether this specific relationship
is satisfied.

We could not find obvious modifications that will address all these
issues. While it is not clear that those cannot be found, this suggests that the simple versions of these models might not be
enough.  A simple extension of  a wide shell $d \gg 1$ or several
separated shells might resolve some of the issues and it might be
essential for the sub-jet model allowing moderate values of the sub-jet's
Lorentz factor.

This research is supported by the ISF center of excellence in  High Energy Astrophysics (TP \& AL),  Marie Curie  IRG grant (EN),  advanced ERC excellence award and  the Schwartzmann chair (TP).

\end{document}